\documentclass[lettersize,journal]{IEEEtran}
\usepackage{amsmath,amsfonts}
\usepackage{array}
\usepackage[caption=false,font=normalsize,labelfont=sf,textfont=sf]{subfig}
\usepackage{textcomp}
\usepackage{stfloats}
\usepackage{url}
\usepackage{graphicx}
\usepackage{cite}
\usepackage{booktabs}
\usepackage{tabularx}
\usepackage[table]{xcolor}
\newif\ifmarked
\markedfalse
\ifmarked
\newenvironment{rev}{\color{blue}}{}
\else
\newenvironment{rev}{}{}
\fi
\newcommand{\revcolor}{\ifmarked\color{blue}\fi}
\begin{document}

\title{Context Graph: Explainable Multi-Agent \\Decision Chains in AI-RAN}

\author{\IEEEauthorblockN{
    Xijun~Wang,
    Zhaoyang Liu,
    Xiang~Chen,
    Howard~H.~Yang,
    and Chenyuan Feng}
\thanks{Xijun~Wang, Zhaoyang Liu, and Xiang~Chen are with Sun Yat-sen University, China; Howard~H.~Yang is with Zhejiang University, China; and Chenyuan Feng is with University of Exeter, U.K.}
\vspace{-1.5em} }

\maketitle

\begin{abstract}
AI-RAN distributes network control across multiple autonomous agents, namely xApps and rApps operating over shared radio resources. When an alert or degradation appears, operators currently lack a structured way to trace which decision caused it, what context triggered that decision, and how other agents interacted. Existing explainability tools focus on single-model feature attribution or linear event logging, neither of which captures cross-agent decision chains. We propose the \emph{Context Graph}, a heterogeneous directed graph that organizes policies, measurements, decisions, and alerts into a queryable explanation layer. Its foundational primitive is the Decision-Event node acting as a causal junction, where incoming edges capture triggering context and outgoing edges track downstream consequences that are progressively upgraded through counterfactual verification. Instantiated over existing E2, A1, and O1 data flows, our proof-of-concept demonstrates that compliance and diagnosis queries are resolved at sub-millisecond latency over graphs of tens of thousands of nodes, while reliably distinguishing true causal impacts from spurious correlations.
\end{abstract}

\begin{IEEEkeywords}
O-RAN, AI-RAN, explainable AI, multi-agent systems, context graph, causal inference, RIC, xApp.
\end{IEEEkeywords}

\section{Introduction}
The radio access network (RAN) is becoming an ecosystem of autonomous agents \cite{ref_nativeai}. In the Open RAN (O-RAN) architecture, network intelligence is no longer concentrated in a single optimizer. It is distributed across xApps and rApps that observe, decide, and act in parallel, each tuning transmit power, allocating physical resource blocks (PRBs), steering handovers, or adjusting slicing policies over shared radio resources. Recent AI-RAN studies make this abstraction explicit, treating control applications as learned or intent-driven agents embedded across RAN control loops \cite{ref_agentic,ref_agentran,ref_yang}. This multi-agent paradigm promises faster adaptation and finer-grained optimization. However, it also creates a new operational risk. When many learned agents act on overlapping state and coupled radio resources, network behavior reflects collective decisions that no single model accounts for, leaving operators unable to explain why the system acted as it did.

This loss of explainability becomes acute during routine network troubleshooting. For example, an rApp in the Non-Real-Time (Non-RT) RAN Intelligent Controller (RIC) dispatches an A1 policy that raises Ultra-Reliable Low-Latency Communications (URLLC) priority across the network. In the same operating window, a power xApp observes E2 measurements and increases transmit power on one cell. A later E2 report shows a changed signal-to-interference-plus-noise ratio (SINR), and a Virtual Event Streaming (VES) alert over O1 reports a service-level agreement (SLA) violation. The operator must determine whether these events are merely coincidental or causally linked. The raw information exists in policies, measurements, decisions, and alerts, yet no structure links the alert to candidate decisions, triggering measurements, or policy context.

This example exposes a system-level explainability gap. First, existing explainable-AI work in O-RAN has largely focused on model-level attribution, where the goal is to relate a single prediction or decision to its input features \cite{ref_brik,ref_xairan}. Second, AI-RAN frameworks add evidence records, memory, or reasoning traces to make autonomous control more auditable \cite{ref_agentic,ref_agentran}. Yet these traces remain largely linear records of what happened, lacking the structure to express why an action was taken in context or how its consequences branched over time. Third, conflict-management methods come closer to the system-level setting because they model interactions among xApps, but their goal is to detect or mitigate competing actions in the near-real-time loop \cite{ref_graphica,ref_santos,ref_pacifista,ref_wadud}. Similarly, lightweight runtime monitors tracking parameter-KPI dependency shifts can flag that an anomaly occurred, but they cannot identify which decision chain caused it \cite{ref_djidjev}. These approaches are useful within their own scopes, but they still leave the operator's question unresolved. They explain a prediction, record an event sequence, or arbitrate a conflict, yet they do not reconstruct why one decision was taken in a given context, how it interacted with other agents, and what consequences followed.

In this article, we propose the Context Graph as a system-level explanation layer for AI-RAN. Rather than explaining the neural policy inside each xApp, it organizes policies, measurements, agent decisions, and alerts into a queryable decision chain. An operator can then traverse from an alert back to candidate decisions and further back to the policy context and measurements that shaped those decisions. The graph is grounded in existing O-RAN data flows. Its structure is annotated progressively. Edges first record temporal or contextual relations, and they are later upgraded to causal links when counterfactual inference supports an interventional interpretation. With such a layer, explainability shifts from asking what a single agent predicted to tracing how decisions by multiple agents, shaped by policies and measurements, cascaded into the observed network behavior.

\section{The Decision-Chain Explainability Gap}
\label{sec:gap}

\subsection{Coupled Actions and Decision Chains}
As network control is distributed across multiple AI agents, explainability must shift from an isolated model decision to a distributed control process. In AI-RAN, this process is split between Near-Real-Time (Near-RT) RIC xApps that make short-timescale control decisions and Non-RT RIC rApps that optimize policies over longer horizons and dispatch A1 policies downward. Crucially, despite their different objectives and operating timescales, these agents observe overlapping network state and act on coupled radio resources. This coupling makes it harder to attribute system behavior than to explain a single model decision. For example, a power xApp acting on one cell can alter the interference seen by neighboring cells, indirectly affecting the PRB decisions made there. Meanwhile, a Non-RT policy can reshape both agents' priorities before either one acts. The resulting behavior is therefore a decision chain spread across agents, cells, slices, and time.

\subsection{Operational Explainability Needs}
Managing multi-agent RAN operations requires system-level visibility rather than single-model inspection. Operators must be able to query decision chains to understand how coupled actions evolve over time. This need is particularly prominent across three core operational scenarios.

First, during audit and compliance, the operator must verify whether a decision respected the relevant A1 policy, slice priority, and SLA constraints. A decision that appears locally reasonable to an xApp may still violate a policy constraint when evaluated against the global network context.

Second, during fault diagnosis, the operator must identify which sequence of actions led to an observed degradation. When SINR drops or a VES alert reports an SLA violation, the operator needs to trace backward from the physical symptom to candidate decisions, and further backward to the measurements and policies that triggered them.

Third, for policy optimization, the operator must recognize when harmful interaction patterns recur. If two xApps repeatedly create conflicts under similar traffic, load, or slice conditions, an effective explanation must provide a reusable structural representation of the context where the conflict recurs, rather than an isolated feature score.

\subsection{Data Islands and System Opacity}
Currently, operators cannot fulfill these tasks because the required decision chains cannot be reconstructed. However, this limitation does not stem from a lack of raw data. O-RAN already exposes key operational artifacts across its architecture, yet they remain isolated islands. E2 measurements show how cell-level key performance indicators (KPIs) evolve, but they do not reveal which agent acted, which policy was active, or why the observed state changed. Similarly, A1 policies expose operator intent and slice-level constraints, but they do not show how an xApp interpreted that guidance when selecting an action.

A similar disconnect persists after decisions are executed. Although xApp records show that an action was taken, they typically lack the triggering context and downstream consequences needed for a complete audit. At the management plane, O1 VES alerts report symptoms such as SLA violations without identifying the upstream decisions that caused them. Each artifact is informative in isolation, yet none captures the complete path from context to decision to consequence. Because raw evidence remains fragmented across interfaces and timescales, operators lack the structural connections required to trace these distributed pieces as a coherent decision chain.

\subsection{Design Requirements}
The gap above translates into five requirements for system-level explainability in AI-RAN. First, decision traces must be structured, so that policies, measurements, decisions, and alerts can be traversed rather than scanned as a flat sequence. Second, context must be a first-class object, as every operational decision can only be justified against the specific policy and measurement state that triggered it.
Third, the representation must support cross-agent and cross-time traversal. The operator should be able to move from one xApp's action to another agent's related decision, or from an alert to events that occurred several control intervals earlier. Fourth, the relationships connecting events must be progressively causal. The representation should distinguish between mere temporal correlation and verified causality, upgrading the former to the latter through counterfactual verification.
Fifth, the mechanism must be deployable over existing O-RAN data flows. Explanations must be reconstructable from E2 measurements, A1 policies, O1 VES alerts, and decision records without introducing new interfaces or stalling the near-real-time control loop. 

\section{Context Graph Architecture for AI-RAN}
\label{sec:arch}

This section presents the Context Graph architecture. Grounded in four design principles, we detail its integration across O-RAN layers, define the graph schema, and explain how intent-aware query routing and progressive causal annotation operate over this structure.

\subsection{Design Principles}
The architecture is built upon four foundational principles. First, it treats every agent decision as a causal junction. Rather than simply logging actions as flat events, the graph links each decision to its triggering context through incoming edges and to its downstream consequences through outgoing edges. This design transforms the graph into an explanation substrate, enabling each decision junction to capture both why the agent acted and what physical changes followed.

The second principle is progressive causal annotation. Rather than asserting causality upfront, outgoing edges initially capture candidate effects and are upgraded to causal links only after counterfactual verification confirms their dependency. This progressive refinement keeps the graph immediately usable for operational queries while distinguishing temporal correlation from true causality.

The third principle is layered decoupling. Building an explanation layer must never stall the active control loop. The architecture restricts Near-RT components to lightweight, non-blocking telemetry capture on the critical control path, while offloading computationally intensive tasks, including graph assembly and causal inference, to the Non-RT domain. By decoupling fast data capture from asynchronous graph reasoning, the system preserves Near-RT control responsiveness.

The fourth principle is non-intrusive integration. The graph relies on standard O-RAN interfaces, namely E2, A1, and O1, without altering existing control-plane specifications. Instead of modifying core protocols to capture decision provenance, the system ingests agent records via lightweight application-level conventions. This decoupling isolates standard radio telemetry from vendor-specific agent internals, ensuring immediate deployability in real-world networks.

\subsection{Three-Layer Architecture}
\begin{figure*}[!t]
\centering
\includegraphics[width=0.62\textwidth]{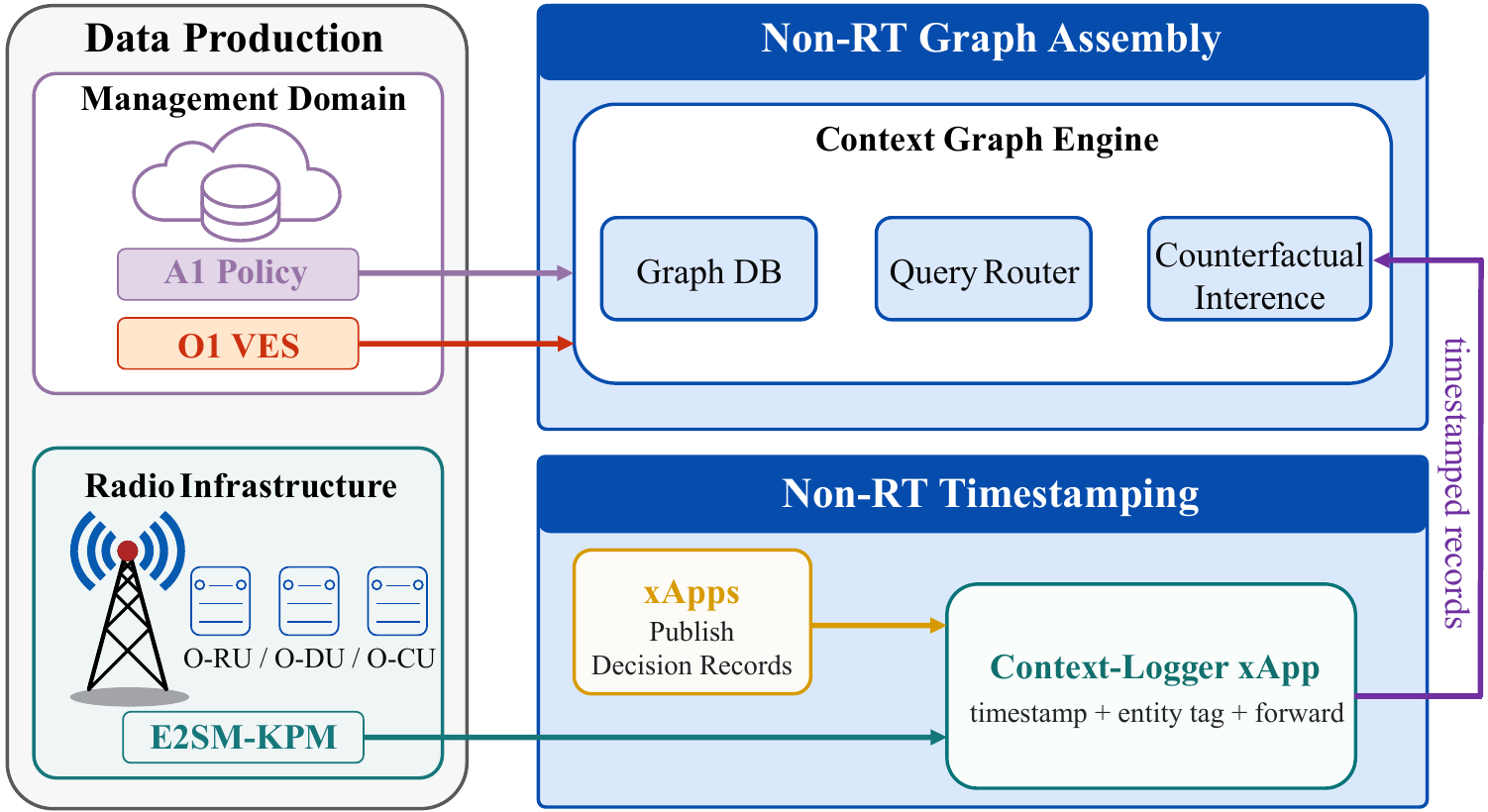}
\caption{Three-layer architecture of the Context Graph in AI-RAN. E2 streams enter through the Context-Logger xApp in the Near-RT RIC for sub-second
   timestamping. A1 policies and O1 VES events are ingested at the Non-RT RIC and SMO. All streams converge in the Context Graph Engine, which
   assembles nodes and edges, runs counterfactual inference, and serves queries.}
\label{fig:arch}
\end{figure*}

The architecture shown in Fig.~\ref{fig:arch} is a hierarchical explanatory framework that spans the existing O-RAN domains. Its primary function is to collect fragmented network events and incrementally assemble them into a unified explanation graph. To achieve this, the system separates event production, Near-RT timestamping, and Non-RT graph assembly across three architectural layers.

At the data-production layer, existing O-RAN components produce the evidence needed for explanation. From the radio infrastructure, the E2 service model for key performance measurement (E2SM-KPM) provides high-resolution measurement reports from the O-RAN radio unit (O-RU), O-RAN distributed unit (O-DU), and O-RAN centralized unit (O-CU). From the management domain, O1 VES events provide asynchronous alarms and anomalies, while A1 exposes operator policy intent and slice constraints. In current deployments, however, these artifacts remain separated by interface and time scale, and the upper layers are therefore needed to preserve their timing and assemble them into one decision chain.

At the Near-RT timestamping layer, a lightweight Context-Logger xApp subscribes to E2 indication messages through the RIC message bus. Other xApps can publish decision records to a designated topic under a lightweight deployment convention. The Context-Logger xApp assigns sub-second timestamps, tags network entities, and forwards the enriched records to the Non-RT engine. \begin{rev}It reads the identifier of an agent from the messages that the agent exchanges with the RIC, and looks up the agent in the deployment records of the RIC platform.\end{rev} If explicit decision logs are unavailable, it forwards standard E2 control records, leaving the Non-RT engine to infer decision nodes from subsequent changes in KPM reports. This stateless forwarding design keeps graph assembly and causal reasoning off the critical control path, while capturing the precise timestamps needed to establish the chronological sequence of events within each control interval.

At the Non-RT graph-assembly layer, the Context Graph Engine handles the computationally heavy tasks. It aggregates timestamped telemetry, xApp decision records, A1 policies, and O1 VES events across the SMO and Non-RT RIC. Over this aggregated data, the engine establishes temporal, semantic, and candidate causal edges, and collaborates with an offline counterfactual inference module to upgrade causal links. On top of this assembled graph, it provides an intent-aware query router that allows operators to trace decision chains on demand. This completes the pipeline, transforming fragmented network evidence into an auditable and queryable explanation layer.

\subsection{Graph Structure}
\begin{figure*}[!t]
\centering
\includegraphics[width=0.8\textwidth]{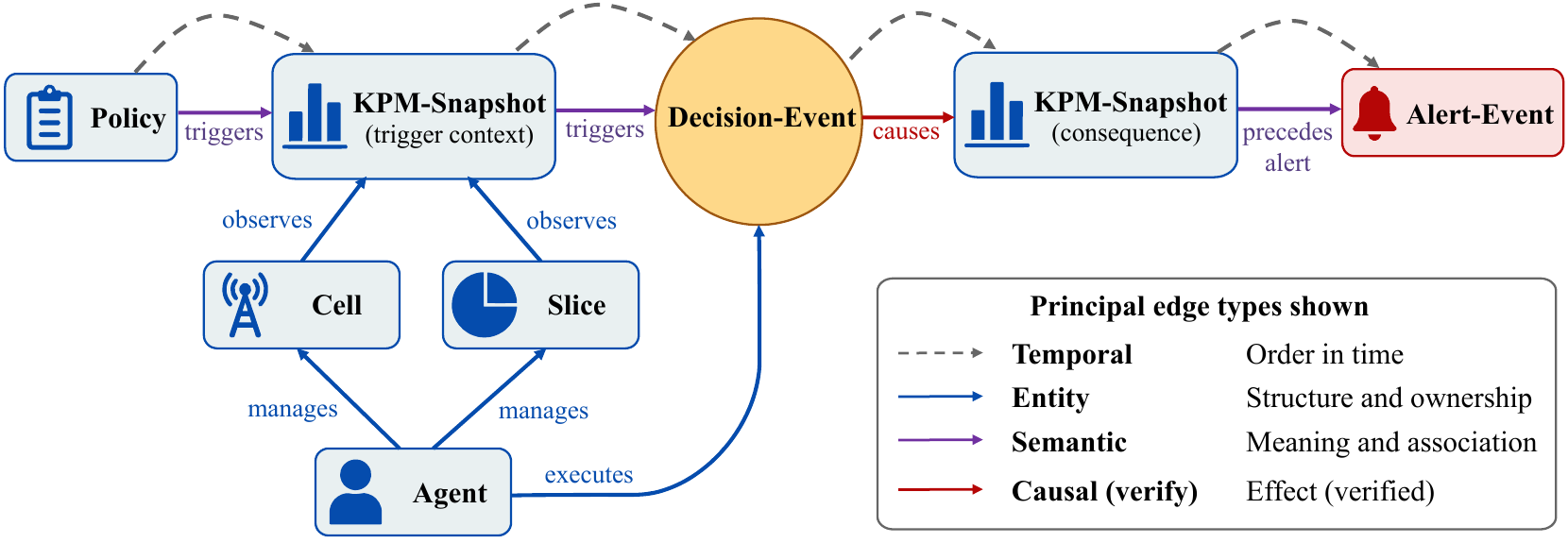}
\caption{Graph structure of the Context Graph, centered on the Decision-Event node as a causal junction. Incoming edges (left) capture the triggering context; outgoing edges (right) capture verified consequences. For clarity, only the upgraded \texttt{causes} edge is shown. Before counterfactual verification, it exists as a \texttt{candidate\_effect} edge.}
\label{fig:contextgraph}
\end{figure*}

To unify fragmented O-RAN telemetry into a coherent topology, the Context Graph is formulated as a heterogeneous directed graph as shown in Fig.~\ref{fig:contextgraph}. It employs a concise yet expressive schema of node and edge types to capture explicit operational relationships among network artifacts. Rather than treating context as an isolated static attribute, the graph models it as an active topological neighborhood around each decision, seamlessly linking triggering conditions to downstream consequences so that complete decision chains can be reconstructed through graph traversal. \begin{rev}This graph serves as a logical model that can be realized either in a native graph database or via relational tables with indexed foreign keys.\end{rev}

As summarized in Table~\ref{tab:nodes}, the graph defines seven node types organized around the Decision-Event node as its core causal junction. Three entity nodes (Cell, Slice, and Agent) ground each decision in its spatial, administrative, and control scope, while three evidence nodes (KPM-Snapshot, Policy, and Alert-Event) capture operational observations and constraints. The Decision-Event node binds these elements into an auditable chain, where incoming edges reflect triggering conditions and outgoing edges represent downstream consequences. \begin{rev}Table~\ref{tab:nodes} distinguishes these node types by their operational sources: standard O-RAN interfaces (E2, A1, and O1), static deployment metadata (for Cell and Agent), and agent provenance (ingested via logging conventions or inferred from E2 control records and subsequent KPM shifts).\end{rev}

\begin{table}[!t]
\caption{Context Graph Node Types\label{tab:nodes}}
\centering
\footnotesize
\renewcommand{\arraystretch}{1.12}
\begin{tabularx}{\columnwidth}{p{1.7cm}Xp{3.0cm}}
\toprule
\textbf{Type} & \textbf{Description} & \textbf{Source} \\
\midrule
\rowcolor{blue!10} KPM-Snapshot & Measurement snapshot of a cell & E2SM-KPM report \\
\rowcolor{blue!10} Policy & Operator intent or SLA constraint & A1 interface \\
\rowcolor{blue!10} Slice & Network slice & A1 policy \\
\rowcolor{blue!10} Alert-Event & SLA violation or anomaly alert & VES over O1 \\
\rowcolor{green!10} Cell & Cell entity & Topology configuration \\
\rowcolor{green!10} Agent & xApp or rApp instance & {Agent identifier in messages, and RIC platform records} \\
\rowcolor{orange!15} Decision-Event & One action taken by an agent & { xApp decision record, or inferred from E2 control records and KPM changes} \\
\bottomrule
\end{tabularx}
\end{table}

Edges connect these nodes into explicit reasoning paths, transforming isolated events into an explanatory structure. As summarized in Table~\ref{tab:edges}, edges are grouped into four relation classes. Temporal edges capture chronological sequence across control intervals. Entity edges associate decisions and telemetry with their corresponding agents, cells, and slices. Semantic edges define operational dependencies, including policy triggers, metric associations, and alert precursors. Finally, causal edges track interventional effects, distinguishing unverified candidate consequences (\texttt{candidate\_effect}) from counterfactually verified causal impacts (\texttt{causes}). Classifying edges into these distinct semantic classes enables the query router to traverse only intent-relevant paths rather than scanning the entire graph.

\begin{table}[!t]
\caption{Context Graph Edge Types\label{tab:edges}}
\centering
\footnotesize
\renewcommand{\arraystretch}{1.12}
\begin{tabularx}{\columnwidth}{p{2.72cm}Xp{1.2cm}}
\toprule
\textbf{Type} & \textbf{Meaning} & \textbf{Relation Class} \\
\midrule
\rowcolor{gray!10} \texttt{temporal} & A occurs before B & Temporal \\
\rowcolor{cyan!10} \texttt{manages} & Agent governs a cell or slice & Entity \\
\rowcolor{cyan!10} \texttt{executes} & Agent performs a decision & Entity \\
\rowcolor{cyan!10} \texttt{observes} & Cell or slice produces a KPM report & Entity \\
\rowcolor{blue!10} \texttt{triggers} & Policy or KPM context triggers a decision & Semantic \\
\rowcolor{blue!10} \texttt{correlates} & KPM changes are associated within the same analysis window & Semantic \\
\rowcolor{blue!10} \texttt{precedes\_alert} & { KPM snapshot precedes an alert and is a candidate explanation} & Semantic \\
\rowcolor{orange!15} \texttt{candidate\_effect} & Physically plausible decision consequence & Causal \\
\rowcolor{orange!15} \texttt{causes} & { Consequence supported by counterfactual verification} & Causal \\
\bottomrule
\end{tabularx}
\end{table}

\subsection{Intent-Aware Query Routing}
A single Context Graph serves diverse explanatory tasks without requiring full-graph traversals. The query router first maps an operator inquiry to the relation classes in Table~\ref{tab:edges}, and then navigates only the edge types aligned with that operational intent. This design adapts the concept of intent-driven memory routing \cite{ref_magma}, tailoring it specifically to multi-agent decision chains in O-RAN.

For compliance, the router starts from a Decision-Event and follows \texttt{triggers} edges back to the active policy and governing KPM context. It then traverses \texttt{executes}, \texttt{manages}, \texttt{observes}, and \texttt{temporal} edges to associate the action with the responsible agent, cell, slice, and measurement window. For diagnosis, the router initiates at a symptom such as a SINR drop or alert, tracing backward along \texttt{precedes\_alert} and causal edges to candidate decisions, verifying which consequence edges have been upgraded to \texttt{causes}. For policy optimization, the router follows \texttt{correlates} and \texttt{temporal} edges across successive control windows to identify recurring conditions under which conflicts or alerts emerge. Because each intent activates only a focused subset of edge types, query latency scales with the explanatory path length rather than the overall graph size.

\begin{rev}
To deliver intuitive explanations without requiring operators to inspect raw graph topologies, query results are formatted into a standardized four-part template. Observed facts present empirical events, including alerts, decisions, policies, and telemetry snapshots. Candidate explanations identify decisions structurally linked to the symptom, marking each as verified or candidate. Model-supported judgments quantify verified \texttt{causes} edges along with their estimated effect sizes and model inputs. Finally, unresolved evidence explicitly highlights relationships that current data or models cannot substantiate, preventing overconfident attributions.
\end{rev}

\subsection{Progressive Causal Annotation}
\label{sec:causal}
Not every consequence edge in the graph carries equal weight. A \texttt{candidate\_effect} edge records a physically plausible link between a decision and a subsequent KPM change, but it does not claim that the action caused that outcome. That claim requires counterfactual verification to upgrade the edge to \texttt{causes}. \begin{rev}A \texttt{causes} edge is thus an attribution supported by model verification, reflecting both model fidelity and input quality.\end{rev} We interpret causality in the framework of Pearl's causal ladder \cite{ref_pearl}, establishing counterfactual dependency where the observed change would not have occurred without the action. While causal methods have recently been applied to RAN configuration recommendation \cite{ref_blinc}, existing studies primarily evaluate static parameter-to-KPI dependencies. In contrast, our approach tracks dynamic, multi-agent decision chains across operational control loops.

Causal annotation proceeds in two phases. In the initialization phase, the engine creates a \texttt{candidate\_effect} edge from each Decision-Event node to every KPM-Snapshot node within its physical reach. A PRB decision can influence only its local cell, whereas a power adjustment impacts both its own cell and neighboring cells via interference coupling. This physical-reachability constraint, derived from network topology and propagation models, prevents the graph from generating spurious candidate edges.

In the verification phase, the engine evaluates counterfactual outcomes for each candidate edge. Using a radio-environment model of the network, it estimates what the KPM trajectory would have been without the decision. When the difference between the observed and counterfactual states exceeds a significance threshold, the edge is asynchronously upgraded to \texttt{causes} and annotated with its quantitative performance impact. \begin{rev}Each decision is evaluated independently. When multiple concurrent actions target the same cell or measurement window, all associated candidate edges remain preserved with their individual verification status, enabling diagnosis queries to expose concurrent influences rather than prematurely enforcing a single root cause.\end{rev}

\section{Illustrative Example}
\label{sec:example}

\subsection{Scenario Setup}
To demonstrate how the Context Graph captures multi-agent interactions, we instantiate it on a synthetic yet standards-aligned AI-RAN scenario. The environment comprises 12 cells in a dense urban layout, serving URLLC and eMBB slices controlled by a power xApp and a PRB xApp. A simulator produces periodic A1 policies, 1\,Hz KPM reports, condition-triggered decisions, and O1 SLA alerts. \begin{rev}Traffic load varies dynamically over time, and reported measurements are subject to noise. Agent actions are coupled through shared radio conditions, because power adjustments alter neighboring interference and shift the operational context observed by the PRB xApp. This setup establishes an inter-agent causal dependency, where an action taken by one agent indirectly triggers and constrains the behavior of another.\end{rev}

\begin{figure}[!t]
\centering
\includegraphics[width=0.9\columnwidth]{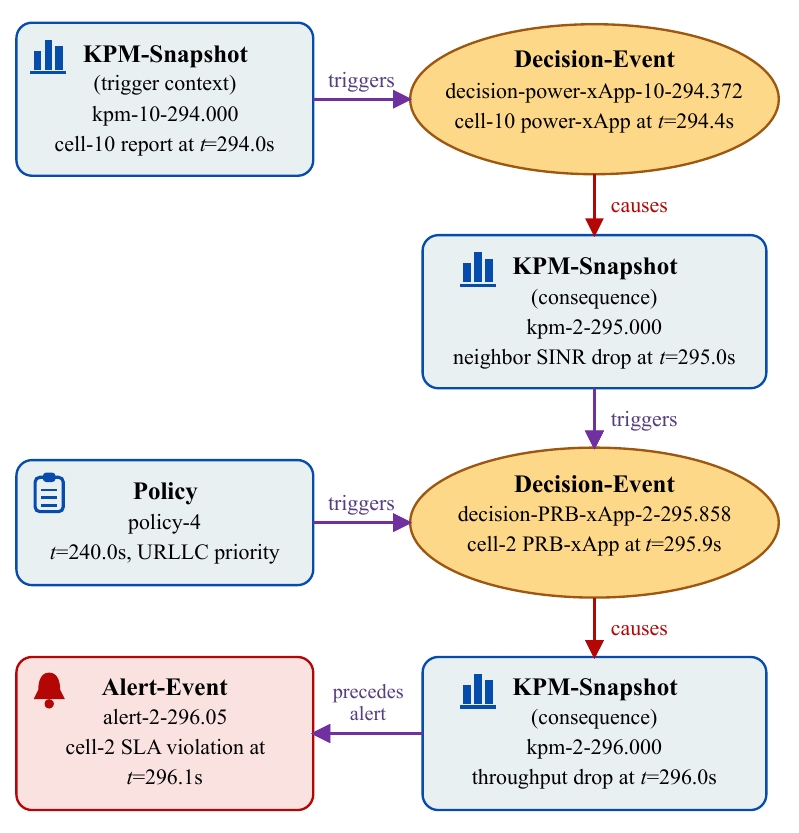}
\caption{Explanatory subgraph for the walkthrough chain. Incoming edges to each Decision-Event node capture the policy and KPM snapshot that triggered the decision, while the highlighted causal edges capture its counterfactually verified consequences. Policy identifiers use a version counter, while KPM, decision, and alert identifiers encode cell index and event time. Agent, cell, and slice entity nodes are omitted to focus on the event-level explanation path.}
\label{fig:casegraph}
\end{figure}

\subsection{The Explanatory Subgraph}
\begin{rev}
The walkthrough example in Fig.~\ref{fig:casegraph} traces an end-to-end decision chain across two interacting agents. At $t=294.0$\,s, a KPM snapshot of cell-10 records high traffic load, prompting the power xApp to raise transmit power at $t=294.4$\,s. This action increases cross-cell interference, lowering the SINR of neighboring cell-2 by 1.2\,dB as captured in its $t=295.0$\,s snapshot. Governed by the active URLLC-priority policy dispatched at $t=240.0$\,s, the PRB xApp on cell-2 reacts at $t=295.9$\,s by shifting resource blocks to the URLLC slice. This reallocation reduces eMBB throughput on cell-2 by 7.1\,Mbps, which is reported at $t=296.0$\,s and promptly triggers an O1 VES alert for SLA violation at $t=296.1$\,s.

A diagnosis query starts from \texttt{alert-2-296.05}, traverses \texttt{precedes\_alert} to \texttt{kpm-2-296.000}, and then follows the \texttt{causes} edge in reverse to reach \texttt{decision-PRB-xApp-2-295.858}. The decision record points to the triggering snapshot \texttt{kpm-2-295.000}, from which another reverse \texttt{causes} edge leads to \texttt{decision-power-xApp-10-294.372}. In four hops, the operator reconstructs the joint chain of the power decision on cell-10 and the PRB decision on cell-2 behind the observed alert. Separately, a compliance query starts from the PRB decision and traverses \texttt{triggers} edges back to \texttt{policy-4} and \texttt{kpm-2-295.000}. Because each query selectively activates only the edge types relevant to its operational intent, the same underlying graph simultaneously supports distinct troubleshooting workflows.

Both queries format their findings using the operator-facing explanation template and fields irrelevant to a given query type remain empty.

\begin{quote}\small
\textit{Diagnosis query}\\
\textbf{Observed facts.} SLA violation on the throughput of cell-2 at $t=296.1$\,s.\\
\textbf{Candidate explanations.} PRB decision of cell-2 at $t=295.9$\,s (verified) and power decision of cell-10 at $t=294.4$\,s (verified).\\
\textbf{Model-supported judgments.} The PRB decision lowered the throughput of cell-2 by 7.1\,Mbps, and the power decision lowered its SINR by 1.2\,dB.
\end{quote}

\begin{quote}\small
\textit{Compliance query}\\
\textbf{Observed facts.} PRB decision of cell-2 at $t=295.9$\,s, taken under the URLLC-priority policy dispatched at $t=240.0$\,s and in response to the KPM snapshot of cell-2 at $t=295.0$\,s.
\end{quote}

\end{rev}

\subsection{Scalability and Causal Evaluation}
\begin{rev}
Table~\ref{tab:scalability} reports graph scale and query latency across five configurations, along with the processing time required for graph assembly, candidate-edge generation, and causal annotation. The first three configurations scale the network across 6, 12, and 24 cells over a 5\,min window, the fourth evaluates 12 cells over 30\,min, and the fifth combines 24 cells with 30\,min. Two trends matter. First, stage execution times scale differently with network load. Increasing cell count from 6 to 24 cells over 5\,min quadruples graph size and assembly cost, while causal annotation rises from 0.018\,s to 0.50\,s. In contrast, extending the observation window from 5 to 30\,min for 12 cells expands the graph six-fold (from 3.8k to 22.9k nodes), and each processing stage scales by a comparable factor. Annotation remains the most computationally demanding stage, reaching 2.0\,s in the 24-cell, 30\,min configuration. Second, query latency remains consistently around 0.005\,ms across all settings when tracing back from alert and KPM nodes with active predecessor chains. This sub-millisecond efficiency is achieved because intent-aware routing traverses only edges matching the query intent rather than the entire graph. Overall, these results demonstrate that the Context Graph maintains low query latency on graphs with up to 45.3k nodes and 96.0k edges.
\end{rev}

\begin{table*}[!t]
\revcolor
\caption{Graph Size, Time of Each Stage, and Query Latency Across Configurations\label{tab:scalability}}
\centering
\footnotesize
\renewcommand{\arraystretch}{1.12}
\begin{tabular*}{\textwidth}{@{\extracolsep{\fill}}lrrrrrrr@{}}
\toprule
\textbf{Config} & \textbf{Nodes} & \textbf{Edges} & \textbf{Assembly (s)} & \textbf{Candidate (s)} & \textbf{Annotation (s)} & \textbf{Compliance (ms)} & \textbf{Diagnosis (ms)} \\
\midrule
6 cells, 5\,min  &  1{,}944 &  4{,}162 & 0.029 & 0.0062 & 0.018 & 0.0051 & 0.0053 \\
12 cells, 5\,min &  3{,}820 &  8{,}216 & 0.053 & 0.010  & 0.080 & 0.0049 & 0.0055 \\
24 cells, 5\,min &  7{,}666 & 16{,}702 & 0.11  & 0.023  & 0.50  & 0.0049 & 0.0054 \\
12 cells, 30\,min & 22{,}883 & 48{,}723 & 0.33  & 0.053  & 0.40  & 0.0049 & 0.0052 \\
24 cells, 30\,min & 45{,}267 & 96{,}002 & 0.76  & 0.11   & 2.0   & 0.0048 & 0.0052 \\
\bottomrule
\end{tabular*}
\end{table*}

\begin{rev}
Beyond scalability, we evaluate whether upgrading a \texttt{candidate\_effect} edge to a \texttt{causes} edge reliably distinguishes true causal effects from spurious correlations. By running counterfactual replays without candidate decisions to establish ground truth across 30 seeds, we observe that the vast majority of temporal candidates do not reflect true causal impact. Relying solely on temporal proximity or empirical KPM fluctuation thresholds proves ineffective, yielding precision below 10\%. In contrast, counterfactual verification achieves 87.3\% precision and 86.6\% recall under typical link-gain estimation error (2\,dB), maintaining over 83\% precision even when traffic telemetry is delayed by 60\,s. Performance degrades gracefully as model mismatch increases, dropping near 50\% only under severe channel estimation error (8\,dB). A \texttt{causes} edge is therefore an attribution grounded in counterfactual verification. When the verification model exhibits error, a false upgrade overstates causal evidence, while a missed upgrade understates it.
\end{rev}

\section{Open Issues and Research Directions}
\label{sec:open}

\subsection{Grounding Causal Claims in Real Deployments}
The reliability of \texttt{causes} edges faces two primary deployment challenges. First, model fidelity governs verification accuracy. \begin{rev}As our evaluation indicates, estimation errors and telemetry delays directly degrade annotation precision and recall.\end{rev} In live networks with rapid fading and proprietary scheduling, an edge labeled \texttt{causes} may reflect simulator bias rather than physical truth. Second, temporal ordering lacks cross-domain guarantees. While the Precision Time Protocol (PTP) synchronizes fronthaul nodes, no standard enforces precise clock alignment between SMO A1 dispatches and Near-RT E2 indications. In sub-second control loops, minor clock skew can invert apparent action-consequence ordering. Bounding both error sources across multi-vendor networks remains a critical open problem.

\subsection{Interface Standardization for Decision Provenance}
While telemetry, policies, and alerts are standardized, xApp decision logging remains proprietary. In the absence of explicit records, the Context-Logger must infer decisions from KPM shifts. \begin{rev}However, post-hoc inference introduces uncertainty in event timing and control magnitudes, potentially causing false upgrades or breaking causal chains entirely. Distinguishing explicit from inferred decisions allows operators to treat inferred traces as provisional evidence.\end{rev}
A key standardization opportunity is defining an E2SM decision-reporting service model. Standardizing how xApps report triggering contexts, actions, and targets closes the provenance gap at source. The research challenge is identifying the minimal schema required for causal verification without burdening latency-sensitive near-real-time loops.

\begin{rev}
\subsection{Graph Lifecycle Management}
In continuous operations, high-frequency periodic KPM telemetry drives unbounded graph expansion, posing a fundamental scalability bottleneck. The open challenge lies in managing graph growth without severing historical decision chains needed for future audits. Simple temporal pruning risks deleting predecessor nodes that anchor active causal paths. Although differentiating nodes by their causal relevance, such as summarizing isolated telemetry while archiving complete decision chains, offers a promising direction, realizing this in practice remains non-trivial. Developing adaptive lifecycle policies that bound graph memory while preserving end-to-end causal traversability over long horizons is an essential topic for future research.
\end{rev}

\subsection{From Explanation to Agent Improvement}
While the Context Graph is currently used for post-hoc diagnosis, its accumulated causal knowledge presents an opportunity to guide agent training. In principle, recurring harmful interaction subgraphs could serve as negative rewards or safety constraints to refine agent policies. However, closing the loop from explanation to multi-agent learning introduces significant theoretical challenges. Because an agent's policy update alters the operating environment of neighboring agents, historical causal traces inevitably suffer from distribution shift. Furthermore, if multiple agents simultaneously adapt their policies based on overlapping causal feedback, the system risks non-convergent oscillations. A fundamental open issue is developing learning mechanisms that extract policy-invariant causal knowledge from graph traces, enabling stable multi-agent co-evolution.

\subsection{Bridging the Graph-to-Operator Gap}
The intent-aware query router assumes that operators can specify query types and starting nodes. In practice, operators naturally troubleshoot using informal inquiries, such as asking why a specific cell suffered an SLA alert. Translating high-level operator intents into precise graph traversal paths presents a significant usability barrier. Coupling the Context Graph with large language model (LLM) interfaces offers a promising solution, where typed graph schemas serve as constrained action spaces to restrict translations to valid topological paths. The open research issue is guaranteeing semantic fidelity during intent translation, preventing generative interfaces from hallucinating non-existent causal dependencies.

\section{Conclusion}
This article introduced the Context Graph to address the system-level explainability gap in multi-agent AI-RAN. \begin{rev}By modeling agent decisions as causal junctions, the architecture transforms fragmented policies, measurements, and alerts into traversable decision chains, progressively upgraded through counterfactual verification.\end{rev} The framework is grounded primarily in standard O-RAN data flows without requiring structural modifications to control-plane protocols. Furthermore, by isolating graph reasoning in the Non-RT layer, the system achieves sub-millisecond query latency and rigorous auditability without compromising active radio control loops. Ultimately, structuring multi-agent interactions into an explicit causal layer provides more than post-hoc troubleshooting. It establishes the foundational trust and provenance substrate necessary for autonomous, self-evolving radio access networks.

\vspace{-1mm}
\vspace{-6mm}
\begin{IEEEbiographynophoto}
{Xijun Wang} (wangxijun@mail.sysu.edu.cn) received the B.S. degree from Xidian University, China, in 2005, and the Ph.D. degree from Tsinghua University, China, in 2012. Currently, he is an Associate Professor at Sun Yat-sen University, China. 
\end{IEEEbiographynophoto}

\vspace{-11mm}
\begin{IEEEbiographynophoto} 
{Zhaoyang Liu} (liuzhy86@mail2.sysu.edu.cn) received the B.E. degree from Sun Yat-sen University, China, in 2022, where he is currently pursuing the Ph.D. degree.
\end{IEEEbiographynophoto}

\vspace{-11mm}
\begin{IEEEbiographynophoto}
{Xiang Chen} (chenxiang@mail.sysu.edu.cn) received the B.E. and Ph.D. degrees from Tsinghua University in 2002 and 2008, respectively. Currently, he is a Full Professor at Sun Yat-sen University, China. 
\end{IEEEbiographynophoto}

\vspace{-11mm}
\begin{IEEEbiographynophoto}
{Howard H. Yang} (haoyang@intl.zju.edu.cn) received the Ph.D. degree from Singapore University of Technology and Design in 2017. Currently, he is an assistant professor at Zhejiang University, China.
\end{IEEEbiographynophoto}

\vspace{-11mm}
\begin{IEEEbiographynophoto}
{Chenyuan Feng} (c.feng@exeter.ac.uk) received the Ph.D. degree from Singapore University of Technology and Design in 2021. Currently, she is a researcher at University of Exeter, U.K and a Marie Skłodowska-Curie Scholar.
\end{IEEEbiographynophoto}

\end{document}